\documentclass[conference]{IEEEtran}
\IEEEoverridecommandlockouts
\usepackage{balance}
\usepackage[hang, marginal]{footmisc}
\usepackage{cite}
\usepackage{amsmath,amssymb,amsfonts}
\usepackage{algorithmic}
\usepackage{graphicx}
\usepackage{textcomp}
\usepackage{xcolor}
\usepackage[draft]{hyperref}

\usepackage[binary-units=true]{siunitx}
\usepackage{tikz}
\usetikzlibrary{positioning, calc, shapes.multipart}
\usepackage{subcaption}
\usepackage{bytefield}
\usepackage{cleveref}

\def\BibTeX{{\rm B\kern-.05em{\sc i\kern-.025em b}\kern-.08em
    T\kern-.1667em\lower.7ex\hbox{E}\kern-.125emX}}
\begin{document}

\title{Latency-aware and -predictable Communication\\ with Open Protocol Stacks for Remote \\Drone Control\\
\thanks{The work is supported by the DFG as part of SPP 1914 ``Cyber-Physical Networking'' under grant HE~2584/4-2.}
}

\author{\IEEEauthorblockN{Marlene Böhmer, Andreas Schmidt, Pablo Gil Pereira and Thorsten Herfet}
\IEEEauthorblockA{\textit{Telecommunications Lab} \\
\textit{Saarland Informatics Campus}\\
Saarbrücken, Germany \\
\{boehmer, andreas.schmidt, gilpereira, herfet\}@cs.uni-saarland.de}
}

\maketitle

\begin{abstract}
In order to create cooperating swarms of Unmanned Autonomous Vehicles (UAVs) that also interact with various other systems and devices, open and free communication systems are mandatory.
This paper presents an implementation of such a communication system to incorporate the Crazyflie nano-drone as a UAV platform.
The protocol stack leverages the open Predictably Reliable Real-time Transport~(PRRT) protocol that adds latency-awareness and -predictability to stacks composed of standard Internet protocols.
To enable the drone to receive and reply to control commands via Wi-Fi, it has been extended with a Raspberry Pi that runs two variants of the Crazybridge---a software to connect the control board to the network.
To evaluate how practical this solution is for the use in control applications, the communication has been analysed with a focus on the latency properties.
Our investigations show that despite using the open protocol stack---and hence opting out of specialised implementations---the resulting latencies are in the same order of magnitude~($4$ to \SI{9}{\milli\second}) as the latency of the proprietary link.
\end{abstract}

\begin{IEEEkeywords}
Communication Architectures, Communication Protocols, Drone Networks, Latency Evaluation
\end{IEEEkeywords}


\section{Introduction}
Dedicated workshops on drones mirror the fact that Unmanned Autonomous Vehicles (UAVs) are gaining attraction and are increasingly used for various application domains~(e.g.\ agriculture~\cite{zorbas:2019:network}, logistics, or emergency services~\cite{McGuire2019}).
In order to support this vast range of applications, the hardware platforms differ greatly, which is also the case for the employed software as well as communication systems~\cite{Hoffmann2004, Spica2013, Manecy2015, Lehnert2013}.
While the former is acceptable, given the different application workloads, the latter must be questioned.
This is in particular the case, because eventually these systems are bound to cooperate and therefore communicate, when they occupy the same areas in the air~\cite{hanscom:2014:unmanned}.
As of today, swarm control is either leveraging identical UAVs or employing middleware that connects different solutions together---two approaches that impede the implementation of applications that incorporate large swarms.
Using a multitude of proprietary and non-interoperable communication platforms is bound to lead to (a)~high costs~(equipping systems with all platforms), (b)~minimal synergies~(components are reinvented and cannot be reused), and (c)~research difficulties~(building widely-usable solutions and appropriate models for closed, heterogeneous systems).
Therefore, it is imperative to use free, open, and interoperable communication systems for building and operating future networks of UAVs.

The Internet architecture~\cite{clark:2018:designing}, with its focus on \emph{interoperability}, has provided the ground for the digitally connected world we live in today.
Even though Internet technologies~(e.g.\ Ethernet or TCP) have been technically inferior\footnote{For instance, having to cope with higher loss rates due to statistical multiplexing and increased latency caused by in-network buffers.} to other solutions present at the time of their invention, Internet technologies have now achieved market domination thanks to their openness and the possibility to produce them in large quantities for competitive prices.
Looking at the expectations of how many UAVs are going to be operated in the next decade~\cite{hanscom:2014:unmanned}, it is evident that the same technology policy should be employed to support these high numbers of devices.

Nevertheless, these Internet protocols often provide lower performance than their proprietary counterparts.
However, various projects aim at improving this performance while maintaining the interoperability, e.g.\ in industrial manufacturing~\cite{pop:2018:enabling}.
In addition to these activities on the lower layers of the ISO/OSI communication model, there is a demand for predictable latency at the transport layer~\cite{gettys:2012:bufferbloat}.
One solution is the \emph{Predictably Reliable Real-time Transport}~(PRRT) protocol~\cite{schmidt:2019:cross-layer},
which is also used in this paper.
We leverage this protocol, together with other off-the-shelf networking and software components, to remote control a nano-drone, i.e.\ the Bitcraze Crazyflie\footnote{\url{https://www.bitcraze.io/portals/research/}}.

The contributions of this paper are threefold:
\begin{itemize}
  \item We describe the design and implementation of a communication path that allows to remote control the Crazyflie drone platform using open protocol stacks.
  \item We show how to leverage the PRRT protocol for this task---an open, latency-aware and -predictable solution.
  \item We provide a thorough latency evaluation of this open system we have developed to show its general suitability for control tasks.
\end{itemize}

The rest of this paper is structured as follows:
\Cref{sec:background} gives further background on UAVs, the Crazyflie platform, as well as the PRRT protocol.
A description of our concrete implementation is given in \Cref{sec:system} and evaluated in \Cref{sec:evaluation}.
Finally, we review related work in \Cref{sec:related-work} and give an outlook in \Cref{sec:conclusion}.


\section{Background}
\label{sec:background}

As UAVs provide an unprecedented tool, due to their special characteristics~(i.e.\ flight dynamics, size and weight, computational power), many different application domains consider to leverage these.
Operating these vehicles efficiently and safely demands a proper cooperation between the cyber components~(i.e.\ information processing) and the physical components~(i.e.\ mechanical flight systems).
Therefore, we consider UAVs~(or drones) as an incarnation of \emph{Cyber-Physical Systems}~(CPS), where control and communication must cooperate to successfully operate in the physical world.
In this realm, a particular challenge is to have appropriate timing and provide predictably low latency for processing as well as communication~\cite{lee:2008:cyber}.

\subsection{The Crazyflie UAV Platform}
For our implementation and evaluation, we use the Bitcraze Crazyflie platform that uses software developed and maintained in the open, i.e.\ as a set of public GitHub repositories\footnote{\url{https://github.com/bitcraze}}.
This platform provides high flexibility, as it can host various \emph{decks} for different navigation modes~(e.g.\ relative / absolute positioning) as well as tasks~(e.g.\ video capture and AI inference).
Furthermore, it is safe to operate this platform in a research and development context, as it is lightweight~(\SI{27}{\gram} for the stock version) but at the same time robust.
In addition to the flying platform itself, the Crazyflie has an ecosystem that includes various tools and libraries that makes it easily accessible for hobbyists, as well as developers from industry and academia alike.
This platform has been used for several applications, in particular research projects in the control engineering domain, e.g.\ autonomous room exploration~\cite{McGuire2019}, swarm orchestration~\cite{Preiss2017}, vision-based control~\cite{palossi:2019:deep-learning, kang:2019:deep-learning}, and remote path optimization~\cite{Araki2017}.

\subsection{Communication Architectures for Networks of UAVs}
UAVs, as well as other CPSs, demand \emph{latency-aware and latency-predictable communication} to maintain the stability of the physical process under control.
Concretely, this means achieving a stable flight while following the desired path as close as possible or being at the correct relative position with regards to other UAVs in a swarm.
If we look at existing communication solutions, we can coarsely categorise them into (a)~\emph{proprietary protocols} that often impede interoperability, and (b)~\emph{open protocols} where interoperability is a major design goal~(as, e.g., in the Internet architecture).

Proprietary communication solutions are typically designed with latency constraints and concrete operating environments in mind, thereby providing tailored performance to the desired application.
This approach leads to the best performance, but comes with a higher price tag~(mass production might not be feasible) and a lack of interoperability~(approach is special-purpose).
The proprietary communication solution for the Crazyflie used in this paper is the \textit{Enhanced Shock Burst}~(ESB) protocol developed by Nordic Semiconductor, which works at the 2.4GHz frequency band and provides the necessary encapsulation for data transmission, as well as an acknowledgement mechanism.
The Crazyflie implements a reliable system setup on the application layer by retransmitting a packet until an acknowledgement for that packet is received, and regularly updates flight commands.
Given its minimalistic design, the underlying ESB could be replaced by standardised physical layers providing a similar service (e.g.\ Bluetooth, Wi-Fi) without a significant impact on the overall communication performance, although this might come at the cost of higher power consumption.

Open solutions make fewer assumptions about the operating environment as well as the application, so that trade-offs are made in favour of higher flexibility and support for a broader range of use cases.
This approach typically sacrifices performance, as it often uses layered design where information is not shared across layers to maximise encapsulation.
If we consider future networks of UAVs, interoperability is necessary to allow successful coordination in the air---at least to ensure safety via coordinated collision avoidance.
As this also demands a certain level of performance with respect to predictably low latency, this paper investigates how Internet protocols can be used to control UAVs.

\subsection{Leveraging Internet Protocols for Remote Control}
In order to establish networked communication between UAVs and ground stations, a straightforward communication stack~(following the OSI model) would use Wi-Fi, IP as well as TCP. This stack provides wireless, networked, process-to-process communication as well as several layers of error control~(multiple checksums, Wi-Fi and TCP retransmissions).
Nevertheless, the 5G New Radio is envisioned to replace the Wi-Fi protocols in some UAV deployments due to its better performance in terms of latency and throughput.
Although 5G has not been considered in this paper, we expect that the proposed solutions would also benefit from the performance increase, since these solutions lie within the upper layers of the stack.
Therefore, exchanging Wi-Fi by 5G New Radio would also benefit the solutions presented in this paper.

With this protocol stack it is possible to leverage existing solutions for wireless communication and allow interconnection with the pre-existing infrastructure, i.e.\ edge computing nodes.
A major concern in this scenario is TCP, which provides full reliability but at the cost of unconstrained and uncontrollable latency~\cite{schmidt:2019:cross-layer}.
Replacing TCP with UDP does not help either, as UAV communication networks demand congestion, flow, and rate control to share communication resources appropriately and avoid increased latency and losses.
Hence, it is necessary to find an alternative on the transport layer that provides better latency characteristics but is an open solution at the same time.

One possible alternative is PRRT, an open transport protocol that provides a latency-aware, partially reliable, in-order datagram delivery service~\cite{schmidt:2019:cross-layer}.
PRRT has several unique mechanisms and APIs that differenciate it from other transport layer protocols.
First, PRRT employs cross-layer pacing to achieve predictably low end-to-end delay by keeping buffers empty.
Second, \emph{Adaptive Hybrid Automated Repeat reQuest}~(AHARQ) is used to adapt the redundancy to the current channel characteristics and application requirements~\cite{schmidt:2019:cross-layer}.
Finally, PRRT's \texttt{send()} and \texttt{recv()} methods are specifically tailored towards the use in control applications, as they allow to (a)~delay the sending for a duration that is appropriate for the current bottleneck of the system and (b)~control the treatment of packet expiry dates upon reception.


\section{System Design and Implementation}
\label{sec:system}
Off-the-shelf, the Crazyflie allows wireless remote control via the proprietary ESB protocol or Bluetooth.
Both have drawbacks in terms of interoperability, because they are either (a)~special purpose~(proprietary) or (b)~only allow small numbers of devices in a short-range network~(Bluetooth).
Therefore, the Crazyflie has been extended with a Raspberry Pi Zero W\footnote{\url{https://www.raspberrypi.org/products/raspberry-pi-zero-w/}} that provides Wi-Fi connectivity~(IEEE 802.11)---allowing us to design and implement an evaluation platform that communicates via open and standardised networking protocols.

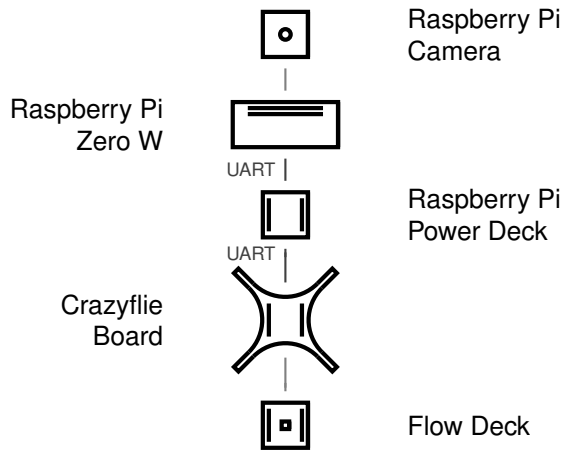
\begin{figure}[t]
  \centering
  \begin{tikzpicture}
\sffamily

\node[align=left, anchor=west] at (1.5,0)	{
	Flow Deck
};
\node (FLOW)	at (0,0)	{
	\begin{tikzpicture}
	\draw[ultra thick] (0.3,0.3) -- (0.3,-0.3) -- (-0.3,-0.3) -- (-0.3,0.3) -- (0.3,0.3) -- (0.3,-0.3);
	\draw[ultra thick] (0.22,0.22) -- (0.22,-0.22);
	\draw[ultra thick] (-0.22,0.22) -- (-0.22,-0.22);
	\draw[ultra thick] (0.05,0.05) -- (0.05,-0.05) -- (-0.05,-0.05) -- (-0.05,0.05) -- (0.05,0.05) -- (0.05,-0.05);
	\end{tikzpicture}
};
\node[align=right, anchor=east] at (-1.5,1.4)	{
	Crazyflie\\Board
};
\node (CF)	at (0,1.4)	{
	\begin{tikzpicture}
	\draw[ultra thick] (0.424,0.484) -- (0.624,0.684) -- (0.684,0.624) -- (0.484,0.424);
	\draw[ultra thick] (-0.424,-0.484) -- (-0.624,-0.684) -- (-0.684,-0.624) -- (-0.484,-0.424);
	\draw[ultra thick] (-0.424,0.484) -- (-0.624,0.684) -- (-0.684,0.624) -- (-0.484,0.424);
	\draw[ultra thick] (0.424,-0.484) -- (0.624,-0.684) -- (0.684,-0.624) -- (0.484,-0.424);
	\draw[ultra thick] (0.424,0.484) arc (135:45:-0.6);
	\draw[ultra thick] (0.484,0.424) arc (135:225:0.6);
	\draw[ultra thick] (-0.424,-0.484) arc (135:45:0.6);
	\draw[ultra thick] (-0.484,-0.424) arc (135:225:-0.6);
	\draw[ultra thick] (0.22,0.22) -- (0.22,-0.22);
	\draw[ultra thick] (-0.22,0.22) -- (-0.22,-0.22);
	\end{tikzpicture}
};
\node[align=left, anchor=west] at (1.5,2.8)	{
	Raspberry Pi\\Power Deck
};
\node (POWER)	at (0,2.8)	{
	\begin{tikzpicture}
	\draw[ultra thick] (0.3,0.3) -- (0.3,-0.3) -- (-0.3,-0.3) -- (-0.3,0.3) -- (0.3,0.3) -- (0.3,-0.3);
	\draw[ultra thick] (0.22,0.22) -- (0.22,-0.22);
	\draw[ultra thick] (-0.22,0.22) -- (-0.22,-0.22);
	\end{tikzpicture}
};
\node[align=right, anchor=east] at (-1.5,4)	{
	Raspberry Pi\\Zero W
};
\node (PI)	at (0,4)	{
	\begin{tikzpicture}
	\draw[ultra thick] (0.7,0.3) -- (0.7,-0.3) -- (-0.7,-0.3) -- (-0.7,0.3) -- (0.7,0.3) -- (0.7,-0.3);
	\draw[ultra thick] (-0.5,0.22) -- (0.5,0.22);
	\draw[ultra thick] (-0.5,0.14) -- (0.5,0.14);
	\end{tikzpicture}
};
\node[align=left, anchor=west] at (1.5,5.2)	{
	Raspberry Pi\\Camera
};
\node (CAMERA)	at (0,5.2)	{
	\begin{tikzpicture}
	\draw[ultra thick] (0.3,0.3) -- (0.3,-0.3) -- (-0.3,-0.3) -- (-0.3,0.3) -- (0.3,0.3) -- (0.3,-0.3);
	\draw[ultra thick] (0,0) circle (0.07);
	\end{tikzpicture}
};

\draw[-, thick, gray] (PI) -- (CAMERA);
\draw[-, thick, darkgray] (POWER) -- node[left] {\scriptsize UART} ++ (PI);
\draw[-, thick, darkgray] (CF) -- node[left] {\scriptsize UART} ++ (POWER);
\draw[-, thick, darkgray] (0,1.9) -- (0,2.3);
\draw[-, thick, gray] (FLOW) -- (CF);
\draw[-, thick, gray] (0,0.9) -- (0,0.5);
\end{tikzpicture}
  \caption{Overview of the Crazyflie Hardware components.}
  \label{graphic/cf-hardware-overview}
\end{figure}

\subsection{System}
\label{sec:system-drone}
The key hardware component of our drone is the Bitcraze Crazyflie~2.1\footnote{\url{https://www.bitcraze.io/crazyflie-2-1/}}, which is a drone control board containing two microcontrollers and essential sensors~(e.g.\ accelerometer and pressure sensor).
Apart from the control board, the drone hosts several other components as seen in \Cref{graphic/cf-hardware-overview}.
Below the Crazyflie control board, the Bitcraze FlowDeck~v2\footnote{\url{https://www.bitcraze.io/flow-deck-v2/}} measures the distance from the ground and horizontal movement to provide basic position estimation.
Above the Crazyflie control board, the Raspberry Pi Zero W is mounted through the Raspberry Pi Power Deck---a prototype module provided by Bitcraze.
A further extension is the Raspberry Pi Camera 2.1 connected to the Raspberry Pi to capture video.

As the additional hardware~(in comparison to the stock Crazyflie) increases weight and size of the drone, a custom 3D-printed frame holds slightly bigger motors (\SI{8.5}{\milli\meter}) with longer propellers (\SI{66}{\milli\meter}).
To have enough power to fly for a reasonable amount of time also a battery with a higher capacity of \SI{500}{\milli\ampere{}\hour} is used.
Nevertheless, the drone and all its hardware components have a weight of only \SI{75}{\gram}.

The Raspberry Pi added to the drone is responsible for enabling the control communication and transmitting video---leveraging its built-in radio chips for Wi-Fi.
For this, the Raspberry Pi is running \emph{Raspbian Buster}~(2019-09-26 release) as an operating system with a \texttt{PREEMPT\_RT} kernel patch.
To further optimise the performance, most services that come with Raspbian by default have been disabled to have a minimal number of services running in parallel to control the drone and transmit video.

\begin{figure*}[t]
	\centering
	\begin{tikzpicture}
\sffamily

\small
\node(n1) {Python script};
\node[below of=n1, yshift=-30, minimum width=1.5cm](n2) {cflib};
\node[right of=n2, align=center, xshift=70](n3) {crazyradio\\firmware};
\node[right of=n3, align=center, xshift=100](n4) {crazyflie\\nrf firmware};
\node[right of=n4, align=center, xshift=70](n5) {crazyflie\\firmware};
\node[above of=n5, yshift=30](n6) {Motors};
\node[below of=n4, align=center, yshift=-35](n7) {crazy\\bridge};

\tiny
\draw[->] (n1) -- node[left, align=right] {High level\\commands} ++ (n2);
\scriptsize
\draw[<->, orange] (n2) -- node[above, align=center, black] {USB} ++ (n3);
\draw[<->, dash dot dot, orange] (n3) -- node[above, align=center, black] {ESB} ++ (n4);
\draw[<->, orange] (n4) -- node[above, align=center, black] {UART} ++ (n5);
\draw[->] (n5) -- (n6);
\draw[<->, dash dot dot, blue] (n2) -- (n2 |- n7) -- node[above, align=center, black] {PRRT} ++ (n7);
\draw[<->, blue] (n5) -- (n7 -| n5) -- node[above, align=center, black] {UART} ++ (n7);

\draw[darkgray] ($(n1.north west)+(-0.1,0.1)$) rectangle ($(n3.south west)+(-1.4,-0.5)$);
\node[darkgray, anchor=west, align=left] at ($(n2.south west)+(-0.4,-0.9)$) {PC};
\draw[darkgray] ($(n3.north west)+(-0.2,0.1)$) rectangle ($(n3.south east)+(0.2,-0.1)$);
\node[darkgray, anchor=west, align=left] at ($(n3.south west)+(-0.1,-0.45)$) {Crazyradio\\USB Dongle};
\draw[darkgray] ($(n4.south west)+(-0.4,-0.5)$) rectangle ($(n6.north east)+(0.5,0.1)$);
\node[darkgray, anchor=west, align=left] at ($(n4.south west)+(-0.3,-0.7)$) {Crazyflie};
\draw[darkgray] ($(n7.north west)+(-0.6,0.1)$) rectangle ($(n7.south east)+(0.6,-0.1)$);
\node[darkgray, anchor=west, align=left] at ($(n7.south west)+(-0.5,-0.45)$) {Raspberry\\Pi Zero W};
\draw[darkgray] ($(n6.north east)+(0.7,0.3)$) rectangle ($(n7.south west)+(-1.05,-0.9)$);
\node[darkgray, anchor=west, align=left] at ($(n7.south west)+(-1,-1.1)$) {Drone};

\draw[gray] ($(n4.north west)+(-0.2,0.1)$) rectangle ($(n4.south east)+(0.2,-0.1)$);
\node[gray, anchor=west, align=left] at ($(n4.south west)+(-0.1,-0.3)$) {Radio chip};
\draw[gray] ($(n5.north west)+(-0.2,0.1)$) rectangle ($(n5.south east)+(0.2,-0.1)$);
\node[gray, anchor=west, align=left] at ($(n5.south west)+(-0.2,-0.3)$) {Control chip};
\end{tikzpicture}
	\caption{Overview of the traditional (orange) and newly implemented (blue) Crazyflie communication paths. Coloured lines depict wired~(solid) and wireless~(dashed) links.}
	\label{graphic/cf-communication-paths}
\end{figure*}
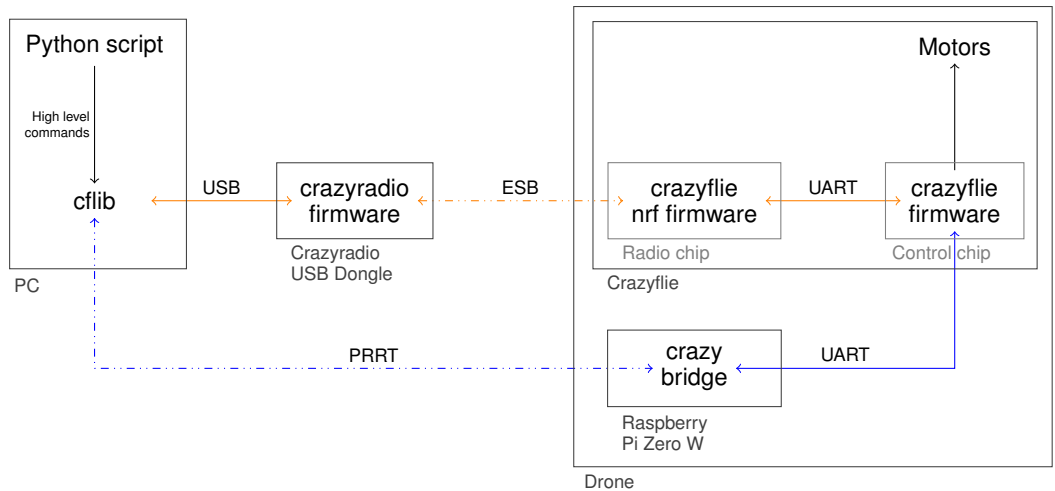

\subsection{Communication Paths}
The Crazyflie is controlled by Bitcraze's application layer protocol called \emph{Crazy Real-Time Protocol}~(CRTP) that encapsulates all control or feedback messages.
These CRTP packets have to be transmitted between the controlling device, e.g.\ a PC, and the Crazyflie.
The traditional communication path, including a proprietary radio link as wireless connection to the drone, and the newly developed communication path that uses open protocols to establish a wireless connection are shown in \Cref{graphic/cf-communication-paths}.

In both cases, a Python script can leverage Bitcraze's Crazyflie Python Library~(\texttt{cflib}) to send control messages.
The traditional radio communication path relies on the Crazyradio USB Dongle to send CRTP messages via the ESB protocol.
Therefore, this path is called \textit{Crazyradio communication path}.
On the drone the ESB messages are received by the microcontroller responsible for radio communication~(Radio chip) and the included CRTP packet is passed on to the microcontroller running the main application~(Control chip) via UART.

The new communication path uses PRRT on the transport layer of the wireless link and is therefore called \textit{PRRT communication path}.
On this path the \texttt{cflib} transmits the CRTP packages to the Raspberry Pi with the help of a Wi-Fi link and IP addressing.
The Raspberry Pi then bridges the CRTP packets to the Control chip on the Crazyflie via UART.

\subsection{Crazybridge}
The central piece of software required for controlling the Crazyflie via the PRRT communication path is the \texttt{crazybridge} that we developed.
In principle, this component is responsible for bridging the PRRT-link~(that covers the Wi-Fi link) to the UART link~(that covers the on-drone communication) on the Raspberry Pi.

\subsubsection{Python Crazybridge}
The first version of the bridge has been developed in Python and leverages Python drivers for UART and PRRT from the \texttt{cflib}\footnote{\url{https://github.com/bitcraze/crazyflie-lib-python}}.
The Crazybridge itself is a Python script that uses two threads that receive messages from the two interfaces by blocking receive calls and send the messages out on the other interface.

The developed UART and PRRT drivers handling the receive and send calls in the \texttt{cflib} have been made available to Bitcraze by several pull requests and are available in their official software on GitHub by now\footnote{\url{https://github.com/bitcraze/crazyflie-lib-python/pull/140}\\\url{https://github.com/bitcraze/crazyflie-lib-python/pull/141}}.
Internally the UART driver builds up on the functionality of the \texttt{pySerial}\footnote{\url{https://pyserial.readthedocs.io/}} package with two threads, one for receiving CRTP packages and one for sending CRTP packages. Between threads data is passed using queues.
The PRRT driver manages sending and receiving without additional, Python-level threads---concurrent reception and sending is encapsulated in the PRRT module that internally leverages several POSIX threads and wait-free inter-process communication.

\subsubsection{Rust Crazybridge}
\label{section/implementation_rust_bridge}
Initial experiments with the Python-based Crazybridge showed that the on-drone communication increased the latency significantly, although it is still acceptable for control scenarios that can work with \SI{20}{\milli\second} RTT~(cf. \Cref{section/eval_ipt_rtt}).
We found that, in particular, the UART communication between Raspberry Pi and Crazyflie is the bottleneck~(cf. \Cref{section/eval_xlap})---although a physical layer analysis revealed that the Crazyflie answers fast, the Python code handling the serial communication is slow.
Hence, we decided to recreate the bridge using the Rust programming language\footnote{\url{https://rust-lang.org}}.
At the time of writing, there is no \texttt{crazyflie-lib-rust}, so we consider this solution to be less \emph{off-the-shelf} than the Python solution, even though we also open-source the code\footnote{\url{https://git.nt.uni-saarland.de/LARN/RNA/crazyflie-lib-rust}\\\url{https://git.nt.uni-saarland.de/LARN/RNA/crazyflie-bridge}}.
Later evaluations~(cf. \Cref{section/eval_ipt_rtt}) show that indeed this implementation achieves far better performance during operation as well as at start-up.

Similar to the Python-based version, the Rust Crazybridge uses threads for passing the packets received from PRRT to UART and vice versa.
For inter-thread communication, the \texttt{crossbeam} crate is used, in particular the multi-producer multi-consumer channels for message passing, i.e.\ moving packets from one link to the other.
The small and custom \texttt{crazyflie-lib-rust} contains a UART driver and the CRTP packet format.
The UART driver builds up on the \texttt{rppal} crate that provides the UART functionality for the Raspberry Pi.
For the PRRT communication a \texttt{prrt} crate\footnote{\url{http://prrt.larn.systems}} has been created with Rust bindings for the C implementation of PRRT.

\subsection{Firmware}
The control chip on the Crazyflie control board is running the Crazyflie firmware\footnote{\url{https://github.com/bitcraze/crazyflie-firmware}} from Bitcraze, which is in turn based on FreeRTOS.
This has been extended by accepting control communication via UART to enable the connection via the Raspberry Pi.
Our addition including packet parsing for the UART interface and changed packet routing is, by now, available in the Bitcraze mainline repository on GitHub due to a pull request\footnote{\url{https://github.com/bitcraze/crazyflie-firmware/pull/558}}.

\subsection{Video Streaming}
In addition to the control communication, which we evaluate in detail in the next section, we have established video communication by streaming the frames captured via the Raspberry Pi camera to the ground station.
For this task, we have leveraged a Gstreamer\footnote{\url{https://gstreamer.freedesktop.org/}} pipeline, which also uses our gst-prrt\footnote{\url{http://gst-prrt.larn.systems}} plugins, thereby enabling the system to stream real-time video of the drone's perspective---simultaneously using the same network stack we also use for controlling the drone.
In order to achieve this, we use \texttt{raspivid} to generate H.264 video with a resolution of 1280x720 pixels, 30 frames per second, and a target bitrate of \SI{1.7}{\mega bps}.
This stream is directly transmitted using the \texttt{prrtsink} and displayed at the receiver where the video is decoded again.


\section{Evaluation}
\label{sec:evaluation}
In order to show how well remote control of UAVs can be implemented, we compare the proprietary solutions with the open solution we have implemented with respect to latency.

\begin{figure}
	\centering
	\begin{subfigure}{\linewidth}
		\centering
		\scalebox{0.7}{
\begin{tikzpicture}
	\node[circle]	(PC)	at (5.5,0)	{
		\begin{tikzpicture}
		\draw[ultra thick, rounded corners=0.5mm] (0,0) -- (0.8,0) -- (0.8,1) -- (-0.8,1) -- (-0.8,0) -- (0.8,0) -- (1,-0.5) -- (-1,-0.5) -- (-0.8,0) -- (0,0);
		\draw[thick, rounded corners=0.5mm] (0,0.1) -- (0.7,0.1) -- (0.7,0.9) -- (-0.7,0.9) -- (-0.7,0.1) -- (0,0.1);
		\draw[ultra thick, rounded corners=0.5mm] (1,-0.25) -- (0.96,-0.15) -- (1.36,-0.15) -- (1.44,-0.35) -- (1.04,-0.35) -- (1,-0.25);
		\draw[ultra thick, rounded corners=0.5mm] (1.4,-0.25) -- (1.5,-0.25) -- (1.5,0.5) -- (1.4,0.5) -- (1.4,-0.25);
		\node[circle]	(Rays)	at (2,0.5)	{
			\begin{tikzpicture} [scale=1.2]
			\draw[ultra thick] (0.1,0.1) arc (225:135:-0.1414);
			\draw[ultra thick] (0.2,0.2) arc (225:135:-0.2828);
			\draw[ultra thick] (0.3,0.3) arc (225:135:-0.4243);
			\end{tikzpicture}
		};
		\end{tikzpicture}
	};
	\node[circle]	(UAV)	at (10,0)	{
		\begin{tikzpicture} [scale=0.7]
		\draw[very thick] (0.5,0.5) circle (4mm);
		\draw[very thick] (-0.5,0.5) circle (4mm);
		\draw[very thick] (-0.5,-0.5) circle (4mm);
		\draw[very thick] (0.5,-0.5) circle (4mm);
		\draw[ultra thick] (0.5,0.5) -- (-0.5,-0.5);
		\draw[ultra thick] (-0.5,0.5) -- (0.5,-0.5);
		\draw[ultra thick] (0.2,0.2) arc (135:45:-0.2828);
		\draw[ultra thick] (0.2,0.2) arc (135:225:0.2828);
		\draw[ultra thick] (-0.2,-0.2) arc (135:45:0.2828);
		\draw[ultra thick] (-0.2,0.2) arc (225:135:-0.2828);
		\end{tikzpicture}
	};
\end{tikzpicture}
}
		\caption{Crazyradio communication path}
		\label{graphic/cf-setup-original}
	\end{subfigure}
	\begin{subfigure}{\linewidth}
		\centering
		\scalebox{0.7}{
\begin{tikzpicture}
	\node[circle]	(PC)	at (0,0)	{
		\begin{tikzpicture}
			\draw[ultra thick, rounded corners=0.5mm] (0,0) -- (0.8,0) -- (0.8,1) -- (-0.8,1) -- (-0.8,0) -- (0.8,0) -- (1,-0.5) -- (-1,-0.5) -- (-0.8,0) -- (0,0);
			\draw[thick, rounded corners=0.5mm] (0,0.1) -- (0.7,0.1) -- (0.7,0.9) -- (-0.7,0.9) -- (-0.7,0.1) -- (0,0.1);
		\end{tikzpicture}
	};
	\node[circle]	(AP)	at (5,0)	{
		\begin{tikzpicture}
			\draw[ultra thick, rounded corners=0.5mm] (0,0) -- (0.7,0) -- (0.9,-0.5) -- (-0.9,-0.5) -- (-0.7,0) -- (0,0);
			\draw[ultra thick, rounded corners=0.5mm] (0.4,0) -- (0.5,0) -- (0.5,0.75) -- (0.4,0.75) -- cycle;
			\draw[ultra thick, rounded corners=0.5mm] (-0.4,0) -- (-0.5,0) -- (-0.5,0.75) -- (-0.4,0.75) -- cycle;
			\node[circle]	(Rays)	at (1.3,0.4)	{
				\begin{tikzpicture} [scale=1.2]
					\draw[ultra thick] (0.1,0.1) arc (225:135:-0.1414);
					\draw[ultra thick] (0.2,0.2) arc (225:135:-0.2828);
					\draw[ultra thick] (0.3,0.3) arc (225:135:-0.4243);
				\end{tikzpicture}
			};
		\end{tikzpicture}
	};
	\node[circle]	(UAV)	at (9,0)	{
		\begin{tikzpicture} [scale=0.7]
			\draw[very thick] (0.5,0.5) circle (4mm);
			\draw[very thick] (-0.5,0.5) circle (4mm);
			\draw[very thick] (-0.5,-0.5) circle (4mm);
			\draw[very thick] (0.5,-0.5) circle (4mm);
			\draw[ultra thick] (0.5,0.5) -- (-0.5,-0.5);
			\draw[ultra thick] (-0.5,0.5) -- (0.5,-0.5);
			\draw[ultra thick] (0.2,0.2) arc (135:45:-0.2828);
			\draw[ultra thick] (0.2,0.2) arc (135:225:0.2828);
			\draw[ultra thick] (-0.2,-0.2) arc (135:45:0.2828);
			\draw[ultra thick] (-0.2,0.2) arc (225:135:-0.2828);
		\end{tikzpicture}
	};
	\draw[-, very thick] (1.25,-0.5) -- (3.4,-0.5);
\end{tikzpicture}
}
		\caption{PRRT communication path}
		\label{graphic/cf-setup-wifi}
	\end{subfigure}
	\caption{Setup for the different Crazyflie communication paths.}
	\label{graphic/cf-setup}
\end{figure}
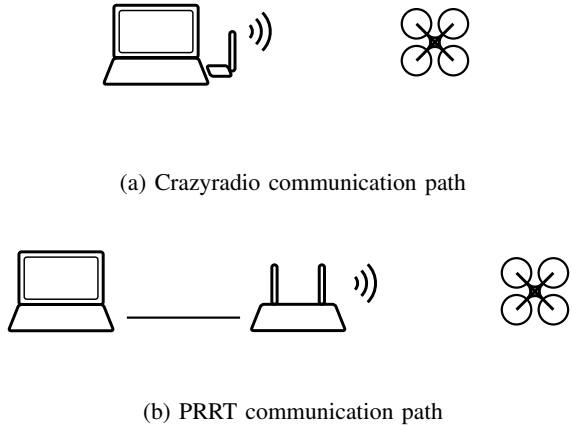

\begin{figure}
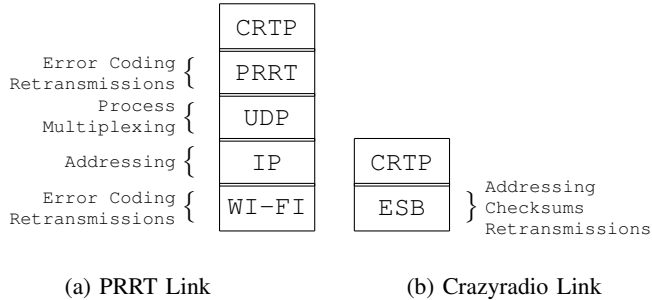

	\centering
	\ttfamily
	\begin{subfigure}[b]{0.45\linewidth}
		\centering
		\begin{bytefield}[bitwidth=0.6em]{6}
			\bitbox{6}{CRTP} \\
			\begin{leftwordgroup}{\begin{minipage}{2.5cm}\scriptsize\raggedleft Error Coding \\ Retransmissions \end{minipage}}
				\bitbox{6}{PRRT}
			\end{leftwordgroup} \\
			\begin{leftwordgroup}{\begin{minipage}{2.5cm}\scriptsize\raggedleft Process \\ Multiplexing \end{minipage}}
				\bitbox{6}{UDP}
			\end{leftwordgroup} \\
			\begin{leftwordgroup}{\begin{minipage}{2.5cm}\scriptsize\raggedleft Addressing \end{minipage}}
				\bitbox{6}{IP}
			\end{leftwordgroup} \\
			\begin{leftwordgroup}{\begin{minipage}{2.5cm}\scriptsize\raggedleft Error Coding \\ Retransmissions \end{minipage}}
				\bitbox{6}{WI-FI}
			\end{leftwordgroup} \\
		\end{bytefield}
		\caption{PRRT Link}
		\label{bytefield/proto-stack-prrt}
	\end{subfigure}
	\hfill
	\begin{subfigure}[b]{0.45\linewidth}
		\centering
		\begin{bytefield}[bitwidth=0.6em]{6}
			\bitbox{6}{CRTP} \\
			\begin{rightwordgroup}{\begin{minipage}{2.5cm}\scriptsize Addressing \\ Checksums \\ Retransmissions \end{minipage}}
				\bitbox{6}{ESB}
			\end{rightwordgroup}\\
		\end{bytefield}
		\caption{Crazyradio Link}
		\label{bytefield/proto-stack-radio}
	\end{subfigure}
	\caption{Underlying protocols used to transmit CRTP packets.}
	\label{bytefield/proto-stack}
\end{figure}

\subsection{System Setup}
For the evaluation of the control communication, the same tests have been executed with both communication paths from \Cref{graphic/cf-communication-paths}.
The setup consists of a drone as described in \Cref{sec:system-drone} and a laptop as the controlling device.
For the traditional radio communication, the Crazyradio dongle is plugged into the USB port of the laptop (\Cref{graphic/cf-setup-original}).
For the PRRT communication path, we have an D-Link DAP-1665 access point connected to the laptop via an Ethernet cable~(\Cref{graphic/cf-setup-wifi}).
The access point is configured to provide 802.11n Wi-Fi in the 2.4GHz band only---the 5GHz band as well as other standards have been disabled.
The reason behind this is that the Raspberry Pi Zero W does not support 5GHz Wi-Fi yet.

On top of the Wi-Fi link, we run standard IPv4 for network addressing, UDP for process multiplexing and PRRT for timing-aware transmission of CRTP packets (\Cref{bytefield/proto-stack-prrt}).
When transmitting CRTP packets via the traditional radio link using the Crazyradio, the protocol stack only consists of ESB as a link layer that provides the fundamental functionality below CRTP~(\Cref{bytefield/proto-stack-radio}).

\subsection{Evaluations}
We evaluate three scenarios of communication with the drone, which are
\begin{description}
  \item [\emph{a}] the Crazyradio communication path using the traditional radio link
  \item [\emph{b}] the PRRT communication path with the Python bridge, and
  \item [\emph{c}] the PRRT communication path with the Rust bridge.
\end{description}
Each scenario has been evaluated multiple times with the generation of a packet log on the controlling laptop.
Two of these executions are given to shed light on the variation of the results~(named 1 and 2 for each of the scenarios).
The evaluations shown in the diagrams relate to the scenarios as follows: \emph{a} radio, \emph{b} prrt python, and \emph{c} prrt rust.

\subsection{Inter-Packet and Round-Trip Latency Analysis}
\label{section/eval_ipt_rtt}

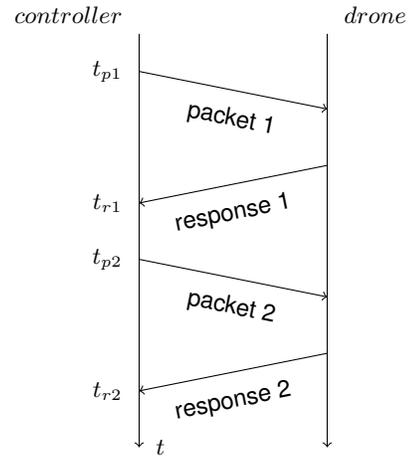
\begin{figure}
	\centering
	\begin{tikzpicture}[my shape/.style={rectangle split, rectangle split parts=#1}, font=\sffamily\small]
\tikzset{input/.style={}}

\def\width{2.5}
\def\height{-5.5}
\def\halfwidth{1.0}

\definecolor{red_color}{rgb} {0.792,0.00,0.125}
\definecolor{blue_color}{rgb} {0.02,0.443,0.69}
\definecolor{grey_color}{rgb} {0.443,0.443,0.443}

\coordinate (left-top) at (0,0);
\coordinate (right-top) at (\width,0);
\coordinate (left-bottom) at (0,\height);
\coordinate (right-bottom) at (\width,\height);

\draw[->] (left-top) -- (left-bottom);
\draw[->] (right-top) -- (right-bottom);

\node [left =1mm of left-top, yshift=0.25cm, my shape=1, rectangle split horizontal] {$controller$};
\node [right =1mm of right-top, yshift=0.25cm, my shape=1, rectangle split horizontal] {$drone$};

\node [right =1mm of left-bottom, my shape=1, rectangle split horizontal] {$t$};

\draw[->] (0,-0.5) coordinate (a)-- node [pos=0.6, sloped, below=0, my shape=2, minimum height=0.75cm, rectangle split horizontal] {packet 1} ++(\width,-0.5) coordinate (b);
\node [left =1mm of a, my shape=1, rectangle split horizontal] {$t_{p1}$};

\draw[<-] (0,-2.25) coordinate (d)-- node [pos=0.55, sloped, below=0, my shape=2, minimum height=0.75cm, rectangle split horizontal] {response 1} ++(\width,0.5) coordinate (c);
\node [left =1mm of d, my shape=1, rectangle split horizontal] {$t_{r1}$};

\draw[->] (0,-3.0) coordinate (e)-- node [pos=0.6, sloped, below=0, my shape=2, minimum height=0.75cm, rectangle split horizontal] {packet 2} ++(\width,-0.5) coordinate (f);
\node [left =1mm of e, my shape=1, rectangle split horizontal] {$t_{p2}$};

\draw[<-] (0,-4.75) coordinate (h)-- node [pos=0.55, sloped, below=0, my shape=2, minimum height=0.75cm, rectangle split horizontal] {response 2} ++(\width,0.5) coordinate (g);
\node [left =1mm of h, my shape=1, rectangle split horizontal] {$t_{r2}$};

\end{tikzpicture}
	\caption{Logging timestamps of control communication.}
	\label{graphic/timeline}
\end{figure}

The packet log taken on the controlling device captures the timestamps of outgoing and incoming packets.
\Cref{graphic/timeline} shows an example of two packet--response pairs with their logged timestamps.
The logs allow to compute inter-packet times~(IPT) and round-trip times~(RTT) to analyse and compare the properties of the communication path.

\begin{figure}[t]
	\includegraphics[width=\columnwidth]{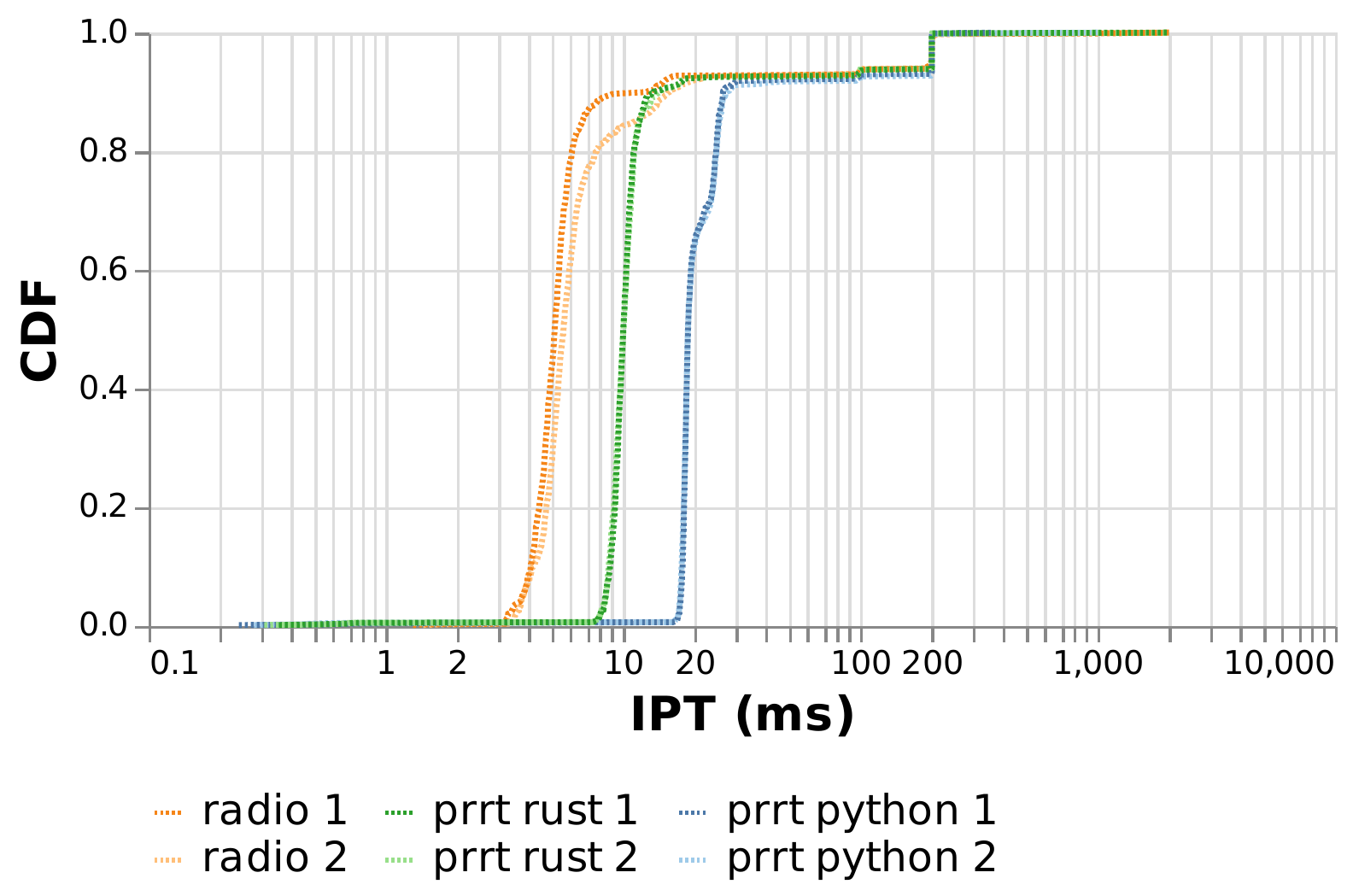}
	\caption{Inter-packet times (IPT) of the control communication.}
	\label{figure/cdf_ipt}
\end{figure}

The IPT reflects how fast packets are transmitted one after each other and it is calculated as the time difference between outgoing packets:
$$\mathrm{IPT}=t_{p2}-t_{p1}$$
\Cref{figure/cdf_ipt} shows the cumulative density functions of the IPT for the three scenarios.
The exact shape of the curve shows several aspects of the system design.
The lower range until \SI{200}{\milli\second} IPT is mainly influenced by the RTT, because packet--response pairs are sent non-pipelined, i.e.\ the system waits for the reply before a new packet request is sent.
The step at \SI{200}{\milli\second} is caused by control command updates that are issued every \SI{200}{\milli\second}.
The rather long tail is due to a time gap between the system setup and the in-flight phase, which causes few packets with high IPT, but this does not indicate a performance issue of the communication path.
As the main difference of the IPTs for the different scenarios is caused by the RTT, we have a further look at this quantity.

\begin{figure}[t]
	\includegraphics[width=\columnwidth]{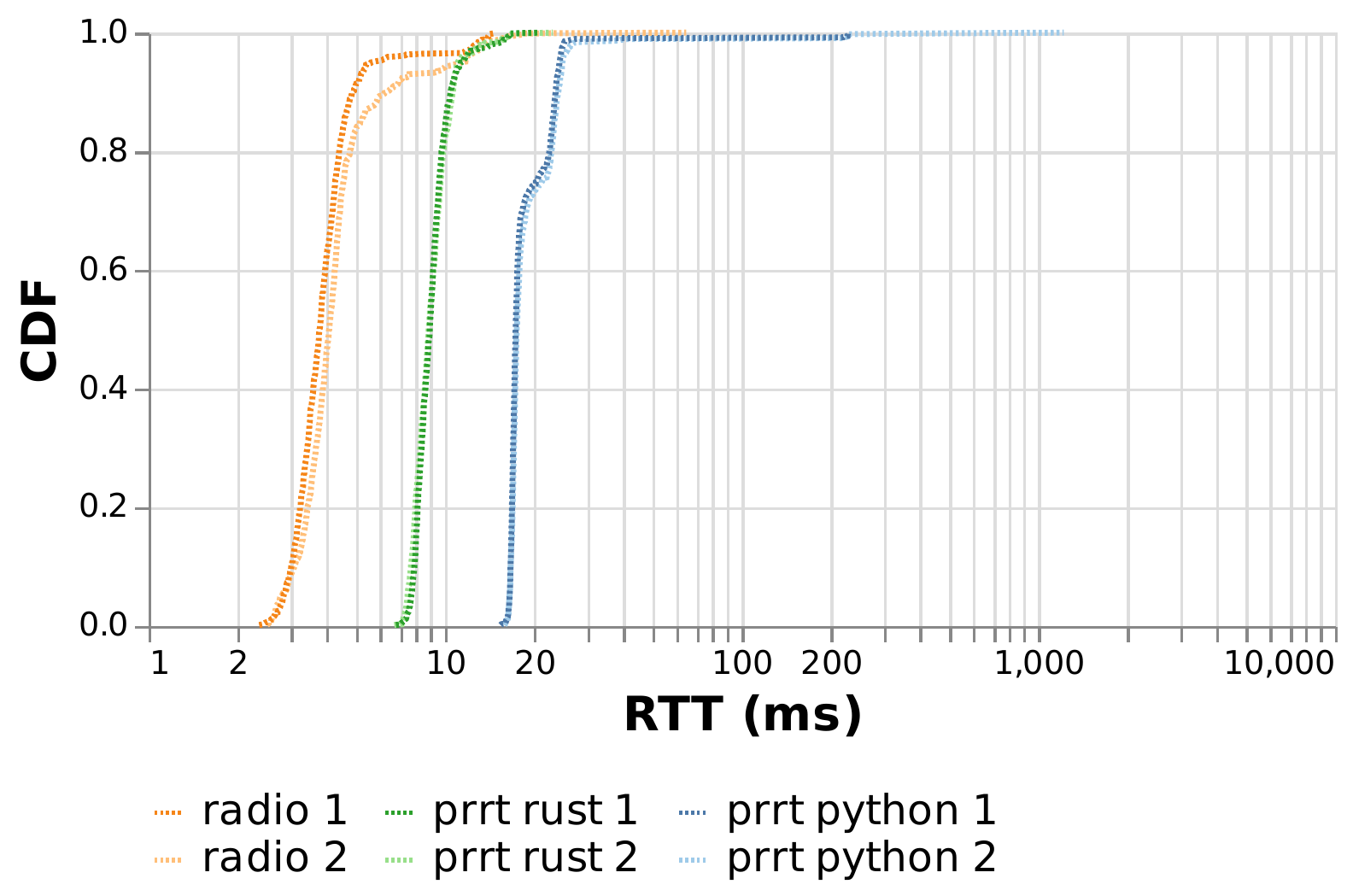}
	\caption{Round-trip times (RTT) for control communication.}
	\label{figure/cdf_rtt}
\end{figure}

The RTT shows the reaction time of the system, i.e.\ it is the latency between sending a request packet ($t_{p1}$) and receiving the corresponding response ($t_{r1}$):
$$\mathrm{RTT}=t_{r1}-t_{p1}$$
\Cref{figure/cdf_rtt} shows the cumulative density functions of the RTT. Here, we only consider packets that receive a response and exclude control commands that are not answered.
If request or response packets are lost, the controller may resend the request packet to get a response.
In such a case the RTT is calculated as the time between the first request packet sent and the first response packet received, which leads to high tail RTTs in the curves in~\Cref{figure/cdf_rtt}.
From the different latencies of the major rise in the curves, it is obvious that the RTT in the Crazyradio communication path is the lowest at about \SI{4}{\milli\second}, followed by the RTT in the PRRT communication path with the Rust bridge at \SI{9}{\milli\second}, and the RTT in the PRRT communication path with the Python bridge at \SI{18}{\milli\second}.
In consequence, exchanging the Python bridge with the Rust bridge significantly improved the performance of the communication path by halving the RTT and approaching the RTT of the Crazyradio communication path.
Even though the RTT of the Rust bridge is still higher than the one for the Crazyradio communication path, this curve is steeper, which means that the RTT is more predictable.
Furthermore, the difference between the two executions for the Crazyradio communication path is higher.
In summary, changing the communication path results in (a)~an increased RTT, which is still well suited for many control applications (even though it may prohibit applications with very high demands), and (b)~an improved predictability of the RTT.

\subsection{Cross-Layer Analysis using X-Lap}
\label{section/eval_xlap}

\begin{figure}[t]
  \includegraphics[width=\columnwidth]{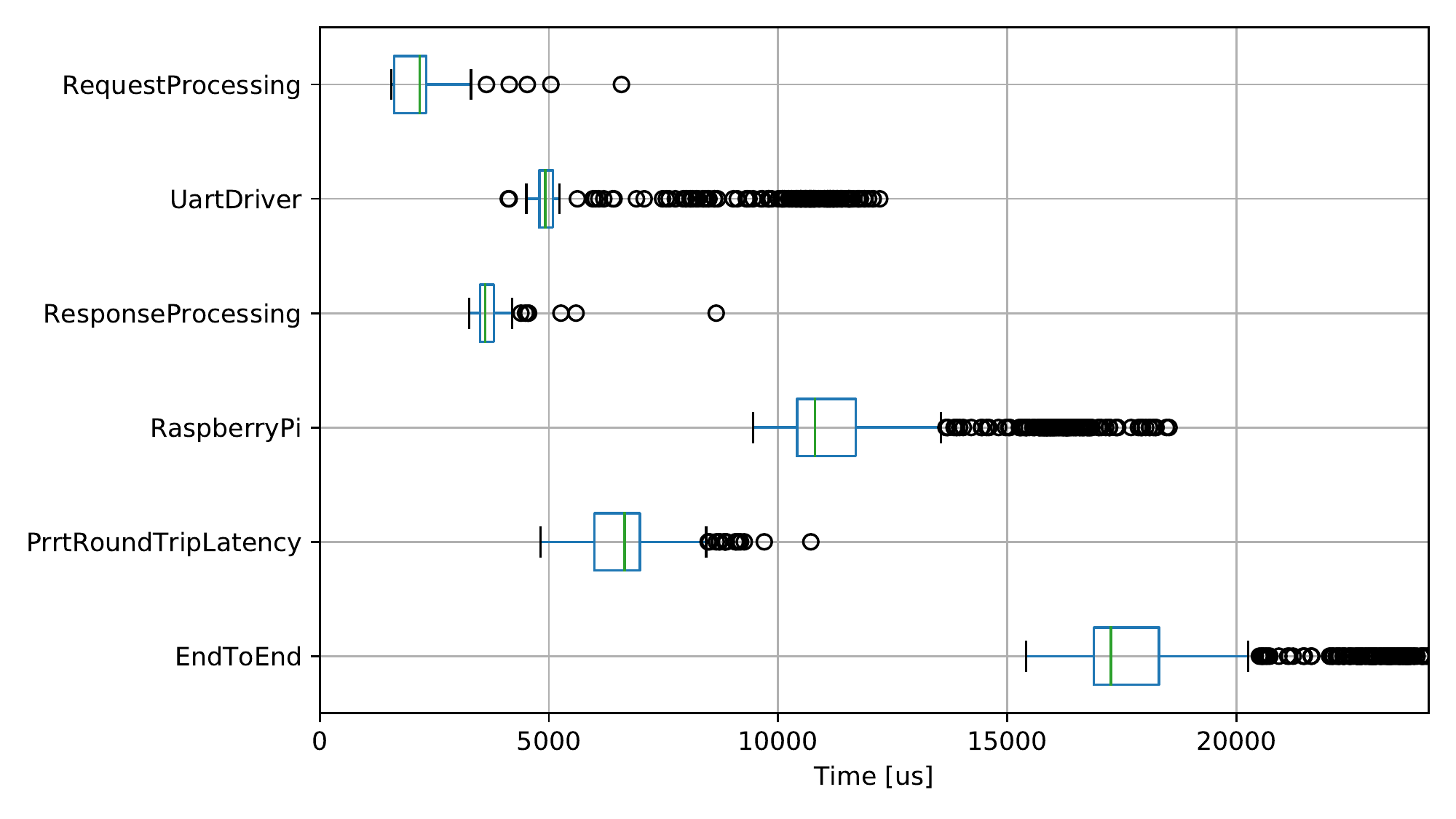}
  \caption{Latencies of different steps in the communication chain for the Python bridge (using run no. 1).}
  \label{figure/python_trace_jitter}
\end{figure}

The primary motivation for exchanging the Python bridge by the Rust bridge was the in-depth analysis of the latency that the PRRT communication path with the Python bridge introduced.
This analysis leverages X-Lap\footnote{\url{https://xlap.larn.systems}}, an open-source cross-layer latency analysis tool we developed previously~\cite{larn:2018:acmsigbed,larn:2019:acmsigbed}.
X-Lap captures traces that allow to deduce how long a packet needed to be processed in a certain step in the communication and processing chain.
This has been applied to the packet--response pairs from the execution number 1 of the PRRT communication path with the Python bridge.
\Cref{figure/python_trace_jitter} shows box plots over packet traces and how these processing and communication latencies vary.

What we can see is that the median end-to-end latency is \SI{17.5}{\milli\second}.
This end-to-end latency is composed of the PRRT round-trip latency---that is the transmission and propagation latencies of the request packet from the controller to the Raspberry Pi together with the same latencies of the response packet from the Raspberry Pi to the controller---and the latency that the Raspberry Pi causes.
The latency of the Raspberry Pi consists of processing the request packet, the latency that the UART driver takes to send the request to the Crazyflie till it gets a response, and the processing of the response packet.

The PRRT round-trip latency has a median latency below \SI{7}{\milli\second}---a result that is inline with \texttt{ping} results of 6 to \SI{7}{\milli\second} we got using the same setup.
As we are using Wi-Fi and IP for the controller to Raspberry Pi communication, this \SI{7}{\milli\second} represent the baseline that we cannot reduce for our experiments\footnote{Using another physical layer, e.g.\ 5GNR, can reduce the latency of the wireless link, but requires different hardware than in our setup.}.
The time spent on the Raspberry Pi with a median of \SI{11}{\milli\second} makes up for about two thirds of the end-to-end latency.
Especially the UART driver takes a lot of time---it accounts for \SI{5}{\milli\second} median latency and up to \SI{12}{\milli\second} in the worst case, and it is a major cause of jitter in this scenario.
The Crazyflie control board answered the requests in the range of \SI{100}{\micro\second} thus the latency must have been introduced on the Raspberry Pi.
These results suggested replacing the software component on the Raspberry Pi, which led us to the development of the Rust bridge~(cf. \Cref{section/implementation_rust_bridge}) to make the latencies smaller and more predictable~(cf. \Cref{figure/cdf_rtt}).

The Rust bridge, which reimplements parts of the Crazyflie Python Library in Rust, does not yet include the logging functionality required for a in-depth latency analysis using X-Lap.
But from the RTT measurements we know that the end-to-end latency is about \SI{9}{\milli\second}.
As the wireless transmission does not differ dependent on the bridge module, the PRRT round-trip latency should be around 6 to \SI{7}{\milli\second}.
Thus, the time usually spent in the Rust bridge is less than \SI{3}{\milli\second}, which is a significant improvement in comparison with the Python bridge with about \SI{11}{\milli\second}.


\section{Related Work}
\label{sec:related-work}

One of the earliest works using open wireless communication for a drone testbed is the STARMAC testbed~\cite{Hoffmann2004}.
The testbed uses the Bluetooth protocol for carrying out multi-agent control tasks~(e.g.\ collision and obstacle avoidance or searching).
As Bluetooth is by design limited to seven devices, we believe that our approach using Wi-Fi, IP, and a real-time transport layer can provide higher scalability.
As they give little details about the software stack, we do not expect that they leverage the same portion of free, open components.
Nevertheless, they provide a very thorough system design and description in their paper.

In the last years, there have been several other works aiming at providing an open-source UAV platform with respect to both hardware and software.
The authors of~\cite{Spica2013} describe an open architecture to operate UAVs, e.g.\ for research and educational purposes.
In contrast to previous work, they employ the \emph{Robot Operating System}~(ROS) and use Wi-Fi standards for transmission---making this solution cheap and easy to reproduce as well as extend.
However, they do not describe details about their networking stack and do not mention how they ensure predictably low latency communication.

In the recent years there has been a trend towards the miniaturization of UAVs, which usually entails the deployment of more constrained devices due to the weight restrictions, what in turn hinders meeting the delay requirements with open operating systems and protocol stacks.
One of such attempts to reduce the UAV weight is the X4-MaG~\cite{Manecy2015}, which in contrast to~\cite{Spica2013} only weights \SI{307}{\gram}.
Their system design is similar to ours, as they also employ a Linux operating system running on the mounted Computer-on-Module chip, as well as leverage Wi-Fi for communication.
Even tough they mention real-time monitoring of the UAV using Wi-Fi, they do not give details about the communication performance.

A further step down the weight scale is $\mu$AV~\cite{Lehnert2013}~(\SI{86}{\gram}), which is also running the open ROS and leveraging Wi-Fi~(802.11b/g) for ground-to-drone communication.
This communication is using the MAVLink message library for efficient payload compression and marshalling---apart from which no special network protocols or algorithms are used, e.g.\ to achieve predictably low latency.

Apart from this related work with a focus on open-source UAV platforms, there is further work on the Crazyflie in particular.

In \emph{Crazyswarm}~\cite{Preiss2017}, a system architecture is described that allows to coordinate 49 Crazyflies in the air, using a motion capturing system for positioning and Crazyradio for communication.
The communication is done using three distinct radio channels and statically group drones into these channels.
They make heavy use of broadcast during flight and request-response messages~(as we do for all commands) for configuration.
Reliability and low latency are achieved by acknowledgements and retransmissions for non-critical unicast messages and a fixed number of proactive transmissions for broadcast messages.
In this respect, our usage of PRRT provides greater flexibility, as we can adapt the hybrid error correction scheme during operation, i.e.\ to make the communication be more proactive or reactive.
Furthermore, our cross-layer pacing allows the systems to adapt their sending rate to the current network bottleneck.

In \cite{McGuire2019}, the authors have also leveraged Crazyradio to do local communication between drones---in the absence of ground infrastructure to coordinate the communication.
As Wi-Fi standards also allow ad-hoc modes of operation~(e.g.\ the 802.11p standard), it could be worthwhile to use our protocol stack for this drone-to-drone communication---leveraging PRRT to achieve latency-awareness and predictability.

The proposal for a network architecture for high volume data collection in agricultural applications~\cite{zorbas:2019:network} depicts a scenario where two radio communication approaches for drones are combined: LoRa and Wi-Fi.
The sensor nodes use LoRa to regularly publish their location~(low information volume) and indicate if data is ready for retrieval.
If this is the case, drones~(acting as data mules) fly near the sensor node and collect the data using Wi-Fi~(large information volume).
For this scenario, our system could assist in making the local Wi-Fi-based communication more flexible in terms of latency and reliability.

The results presented in~\cite{palossi:2019:deep-learning, kang:2019:deep-learning} leverage the images captured by a camera installed on the drones and deep convolutional neural networks~(CNN) to autonomously perform the control function.
Palossi et al.~\cite{palossi:2019:deep-learning} focus on making the drones entirely autonomous, so that the models run on-board without the need for external communication, which is achieved by adding a new device for CNN inference.
In contrast, Kang et al.~\cite{kang:2019:deep-learning} deploy the model in a remote server and transmit the video over the Crazyradio link.
As long as the neural network model is deployed on a server in the edge, our system could provide predictable transport for the transmitted video signal, thus improving the Quality of Control.

There are several related proposals for the lower (i.e.\ physical) layers of communication architectures in drone communication as well as their modelling.
On the technology side, for instance, \cite{hayajneh:2019:coverage} presents backscatter techniques that are used to implement resource-efficient sensor-node-to-drone communication.
On the modelling side, \cite{andryeyev:2019:experimental} focuses on the validation of air-to-ground propagation models and the resulting communication performance.
We consider these as orthogonal to our work, which could leverage these systems and models.


\section{Conclusion}
\label{sec:conclusion}
In order to control large networks of UAVs and let these system cooperate, it is essential to use open communication platforms to maximise interoperability.
These communication platforms must provide latency-awareness and predictably low latency to support the demanding UAV applications.
In this paper, we have shown an open platform~(the Crazyflie) and our open communication stack~(Wi-Fi, IP, PRRT) to remote control the drone.
The evaluation has shown that this stack allows to achieve latencies that are comparable with the ones produced by proprietary communication stacks, but provide a higher potential for interoperability, as open and off-the-shelf solutions have been used.

In future work, we consider to use our platform for evaluating multi-UAV setups to show how open, interoperable platforms can provide benefits over proprietary, special-purpose solution. 
Besides, we plan to leverage PRRT for video-assisted drone control scenarios, which we think allows to further reduce the complexity of drones, since only widely available video encoders are needed instead of tailored AI devices.

\balance
\bibliographystyle{IEEEtran}
\bibliography{larn,references}

\begin{thebibliography}{10}
\providecommand{\url}[1]{#1}
\csname url@samestyle\endcsname
\providecommand{\newblock}{\relax}
\providecommand{\bibinfo}[2]{#2}
\providecommand{\BIBentrySTDinterwordspacing}{\spaceskip=0pt\relax}
\providecommand{\BIBentryALTinterwordstretchfactor}{4}
\providecommand{\BIBentryALTinterwordspacing}{\spaceskip=\fontdimen2\font plus
\BIBentryALTinterwordstretchfactor\fontdimen3\font minus
  \fontdimen4\font\relax}
\providecommand{\BIBforeignlanguage}[2]{{%
\expandafter\ifx\csname l@#1\endcsname\relax
\typeout{** WARNING: IEEEtran.bst: No hyphenation pattern has been}%
\typeout{** loaded for the language `#1'. Using the pattern for}%
\typeout{** the default language instead.}%
\else
\language=\csname l@#1\endcsname
\fi
#2}}
\providecommand{\BIBdecl}{\relax}
\BIBdecl

\bibitem{zorbas:2019:network}
D.~Zorbas and B.~O'Flynn, ``A network architecture for high volume data
  collection in agricultural applications,'' in \emph{Proceedings of the 15th
  International Conference on Distributed Computing in Sensor Systems
  (DCOSS)}.\hskip 1em plus 0.5em minus 0.4em\relax IEEE, 2019, pp. 578--583.

\bibitem{McGuire2019}
K.~N. McGuire, C.~{De Wagter}, K.~Tuyls, H.~J. Kappen, and G.~C. H.~E.
  de~Croon, ``{Minimal navigation solution for a swarm of tiny flying robots to
  explore an unknown environment},'' \emph{Science Robotics}, vol.~4, no.~35,
  2019.

\bibitem{Hoffmann2004}
G.~Hoffmann, D.~G. Rajnarayan, S.~L. Waslander, D.~Dostal, J.~S. Jang, and
  C.~J. Tomlin, ``{The Stanford Testbed of Autonomous Rotorcraft for Multi
  Agent Control (STARMAC)},'' \emph{Proceedings of the 23rd Digital Avionics
  Systems Conference}, vol.~2, 2004.

\bibitem{Spica2013}
R.~Spica, P.~Robuffo~Giordano, M.~Ryll, H.~H. B{\"u}lthoff, and A.~Franchi,
  ``{An Open-Source Hardware/Software Architecture for Quadrotor UAVs},'' in
  \emph{{2nd Workshop on Research, Education and Development of Unmanned Aerial
  System}}, Nov 2013.

\bibitem{Manecy2015}
A.~Manecy, N.~Marchand, F.~Ruffier, and S.~Viollet, ``{X4-MaG: A Low-Cost
  Open-Source Micro-Quadrotor and Its Linux-Based Controller},''
  \emph{International Journal of Micro Air Vehicles}, pp. 89--109, 2015.

\bibitem{Lehnert2013}
C.~Lehnert and P.~Corke, ``{$\mu$AV - Design and implementation of an open
  source micro quadrotor},'' \emph{AC on Robotics and Automation, Eds}, 2013.

\bibitem{hanscom:2014:unmanned}
A.~Hanscom and M.~Bedford, ``{Unmanned Aircraft System (UAS) Service Demand
  2015-2035}.''

\bibitem{clark:2018:designing}
\BIBentryALTinterwordspacing
D.~D. Clark, \emph{Designing an Internet}.\hskip 1em plus 0.5em minus
  0.4em\relax MIT Press, 2018. [Online]. Available:
  \url{https://mitpress.mit.edu/books/designing-internet}
\BIBentrySTDinterwordspacing

\bibitem{pop:2018:enabling}
P.~Pop, M.~L. Raagaard, M.~Gutierrez, and W.~Steiner, ``{Enabling fog computing
  for industrial automation through time-sensitive networking (TSN)},''
  \emph{IEEE Communications Standards Magazine}, vol.~2, no.~2, pp. 55--61,
  2018.

\bibitem{gettys:2012:bufferbloat}
J.~Gettys and K.~Nichols, ``{Bufferbloat: Dark Buffers in the Internet},''
  \emph{Communications of the ACM}, vol.~55, no.~1, pp. 57--65, 2012.

\bibitem{schmidt:2019:cross-layer}
A.~Schmidt, ``Cross-layer latency-aware and -predictable data communication,''
  Ph.D. dissertation, Saarland University, 2019.

\bibitem{lee:2008:cyber}
E.~A. Lee, ``{Cyber Physical Systems: Design Challenges},'' in
  \emph{Proceedings of the 11th IEEE International Symposium on Object and
  Component-Oriented Real-Time Distributed Computing (ISORC)}.\hskip 1em plus
  0.5em minus 0.4em\relax IEEE, 2008, pp. 363--369.

\bibitem{Preiss2017}
J.~A. Preiss, W.~Honig, G.~S. Sukhatme, and N.~Ayanian, ``{Crazyswarm: A large
  nano-quadcopter swarm},'' \emph{Proceedings of the 2017 IEEE International
  Conference on Robotics and Automation (ICRA)}, pp. 3299--3304, May 2017.

\bibitem{palossi:2019:deep-learning}
D.~Palossi, F.~Conti, and L.~Benini, ``{An open source and open hardware deep
  learning-powered visual navigation engine for autonomous nano-UAVs},'' in
  \emph{Proceedings of the 15th International Conference on Distributed
  Computing in Sensor Systems (DCOSS)}.\hskip 1em plus 0.5em minus 0.4em\relax
  IEEE, 2019, pp. 604--611.

\bibitem{kang:2019:deep-learning}
K.~Kang, S.~Belkhale, G.~Kahn, P.~Abbeel, and S.~Levine, ``{Generalization
  through simulation: Integrating simulated and real data into deep
  reinforcement learning for vision-based autonomous flight},'' in \emph{IEEE
  International Conference on Robotics and Automation}, 2019, pp. 6008--6014.

\bibitem{Araki2017}
B.~Araki, J.~Strang, S.~Pohorecky, C.~Qiu, T.~Naegeli, and D.~Rus,
  ``{Multi-robot path planning for a swarm of robots that can both fly and
  drive},'' in \emph{Proceedings of the 2017 IEEE International Conference on
  Robotics and Automation (ICRA)}.\hskip 1em plus 0.5em minus 0.4em\relax IEEE,
  2017, pp. 5575--5582.

\bibitem{larn:2018:acmsigbed}
S.~Reif, A.~Schmidt, T.~H{\"o}nig, T.~Herfet, and W.~Schr{\"o}der-Preikschat,
  ``{X-Lap}: A systems approach for cross-layer profiling and latency analysis
  for cyber-physical networks,'' \emph{ACM SIGBED Review}, vol.~15, no.~3, pp.
  19--24, Aug. 2018, special Issue on 15th International Workshop on Real-Time
  Networks (RTN 2017).

\bibitem{larn:2019:acmsigbed}
------, ``{$\Delta$}elta: Differential energy-efficiency, latency, and timing
  analysis for real-time networks,'' \emph{ACM SIGBED Review}, vol.~16, no.~1,
  pp. 33--38, Feb. 2019, special Issue on 16th International Workshop on
  Real-Time Networks (RTN 2018).

\bibitem{hayajneh:2019:coverage}
A.~Hayajneh, S.~A.~R. Zaidi, M.~Hafeez, D.~McLernon, and M.~Win, ``{Coverage
  Analysis of Drone-Assisted Backscatter Communication for IoT Sensor
  Network},'' in \emph{Proceedings of the 15th International Conference on
  Distributed Computing in Sensor Systems (DCOSS)}.\hskip 1em plus 0.5em minus
  0.4em\relax IEEE, 2019, pp. 584--590.

\bibitem{andryeyev:2019:experimental}
O.~Andryeyev, U.~Onus, V.~Casas, and A.~Mitschele-Thiel, ``{Experimental
  Validation of Air-to-Ground Propagation Models for Low-Altitude Platforms},''
  in \emph{Proceedings of the 15th International Conference on Distributed
  Computing in Sensor Systems (DCOSS)}.\hskip 1em plus 0.5em minus 0.4em\relax
  IEEE, 2019, pp. 591--595.

\end{thebibliography}

\end{document}